\documentclass[runningheads,a4paper]{llncs}

\usepackage{amssymb}
\usepackage{graphicx}
\usepackage[utf8]{inputenc}
\usepackage{lmodern,textcomp}
\usepackage[a4paper=true]{hyperref}
\usepackage[section]{placeins}

\usepackage{url}
\urldef{\mailsa}\path|adrian.paschke@inf.fu.berlin.de|
\urldef{\mailsb}\path|velten@zedat.fu-berlin.de|
\newcommand{\keywords}[1]{\par\addvspace\baselineskip
\noindent\keywordname\enspace\ignorespaces#1}

\renewcommand{\textfloatsep}{10pt}

\begin{document}

\mainmatter  

\title{Semantic Jira  - Semantic Expert Finder in the Bug Tracking Tool Jira}

\titlerunning{Semantic Jira}

\author{Velten Heyn and Adrian Paschke}

\authorrunning{Semantic Jira  - Semantic Expert Finder in the Bug Tracking Tool Jira}

\institute{Corporate Semantic Web, Institute of Computer Science,\\
Koenigin-Luise-Str. 24, 14195 Berlin, Germany\\
\mailsa\\
\mailsb\\
\url{http://www.corporate-semantic-web.de}}

\toctitle{Semantic Jira: }
\tocauthor{Semantic Expert Finder in the Bug Tracking Tool Jira}
\maketitle

\begin{abstract}
The semantic expert recommender extension for the Jira bug tracking system semantically searches for similar tickets in Jira and recommends experts and links to existing organizational (Wiki) knowledge for each ticket. This helps to avoid redundant work and supports the search and collaboration with experts in the project management and maintenance phase based on semantically enriched tickets in Jira.

\end{abstract}

\begin{small}
\section{Introduction}
There is a huge economic potential in the use of Corporate Semantic Web tools in Software Engineering and project management. Bucktracking systems, such as Jira, can benefit from such semantic support. With the information about software bugs, issues, and project tasks cumulated by a bug tracking system and with integrated semantic techniques for transforming this information into meaningful knowledge, it becomes possible to automatically support employees in their daily tasks.

If they need help with a particular task they can browse through similar tickets in the bug tracking system and can reuse the documented solutions to solve their problem. However, with the growing information available in such an enterprise information system it becomes harder to find the relevant tickets.

In this paper we propose a semantic extension to Jira in order to overcome this problem. This Semantic Jira supports semantic search on the knowledge documented in Jira. Similar tickets are semantically inferred by the system and ranked by different metrics (tf-idf, freshness). Furthermore, the most active employees of the found similar tickets are extracted, ranked and proposed as knowledge experts. The underlying rational for this approach is, that they are very likely experts in the field of knowledge, to which this ticket belongs to. The support of asking experts for help has the advantage, that the expert can directly communicate their knowledge and that they can abstract all unnecessary details if the help seeking colleague is not familiar with the domain.

Our expert recommender approach differs from the typical existing solutions which are expert finder search tools. The drawback of searching by users is, that the employees and project managers need to use the right search terms to describe the required skills of an expert for a given problem (bug, issue). This is a non trivial problem, in particular for non-IT persons. In our recommender approach an expert is recommended by a  recommendation system, which uses the reported bug to infer the knowledge field and automatically find the experts in this field. These experts are recommended in the ticket by the Semantic Jira system together with information about similar tasks or tickets (documenting existing problem solutions)  and matching wiki/wikipedia-articles which document existing enterprise knowledge.

The main research questions in the propose solution are:

\begin{enumerate}
\item How to find similar tickets in an automated way? What defines similarity?
\item How do we find possible experts and which information sources should we use for this?
\item Which possibilities do we have to request articles from enterprise Wiki systems which document relevant knowledge?
\end{enumerate}

The benefits of the propose solution,  based on a semantic Jira approach, is that it increases the visibility of implicit and explicit knowledge in the company and that it links this knowledge to the ongoing activities of employees. This helps to avoid redundant work in a company and it helps to optimise the distribution of work on the general resources of employees.

The further paper is structured as follows: Section \ref{relatedwork} describes the related work. Section \ref{concept} introduces the conceptual solution - a Semantic Jira back tracking system.  Section \ref{implementation} gives more details about the implementation of this system. The evaluation in section \ref{evaluation} is based on a user study performed in a company. Finally, section \ref{conclusion} concludes this work.

\section{Related Work}
\label{relatedwork}
“Who Knows about That Bug? Automatic Bug Report Assignment with a Vocabulary-Based Developer Expertise Model“ \cite{Matter} from Dominique Urs Philipp Matter is the largest project we found in our state of art analysis. Their Expert Recommender uses source code analysis to find the most applicable expert for a given bug tracking item. The main drawback of this approach is, that it needs half a year active contribution to the source code from a developer to make proper recommendations. Their use case is to assign developers to new bug tracking items. The fact that they primarily consider source code for expert triage makes the system not usable for non-IT users. That's why we decided to stick to the  data available directly in the bug tracking system.
\\\\
“Expertise Recommender: A Flexible Recommendation System and Architecture” \cite{McDonald:2000} from David W. McDonald and Mark S. Ackerman is another paper which is making use of the change history of source code. A (proprietary) “Tech Support”-Database is used in their field study which was done in a company. With their approach it is only feasible to locate experts within the IT department. The paper further focuses on the architectural aspects of creating a reusable system which can take many different algorithms into account.
\\\\
“Expert Recommender Systems in Practice: Evaluating Semi-automatic Profile Generation” \cite{Reichling:2009} from Tim Reichelt and Volker Wulf is applying another source to tackle this problem. They are using a client program which examines the documents within a folder and subfolder which was selected by the user. This program sends these examined word statistics to the server and compares it with other statistics. It was also considered to use emails as data source but discarded because of privacy protection issues.
\\\\
“Using Domain Ontologies for Finding Experts in Corporate Wikis” \cite{Schafermeier:2011} from Ralph Schäfermeier and Adrian Paschke is taking the wiki entries of users as source. They are using the SEOntology to infer if the users' Wiki contributions are using an experts' knowledge language. They also include an author reputation metric to gain more precision.
\\\\
All these previous works have the problem that they are not sufficiently integrated into the task workflow of an employee which is directly managed  and supported by a bug tracking and project management system such as Jira.

\section{Ticket Recommender and Expertfinder Model}
\label{concept}

Expertise represents the implicit knowledge and the competencies of an expert. The proposed conceptual solution allows inferring expertise from an expertise model which considers the knowledge fields of a ticket saved inside typical bug tracking systems such as Jira.
\\
Typical roles involved in a ticket for such a  model are:

\begin{itemize}
  \item the ticket reporter / creator
  \item the ticket assignee / solver
  \item the ticket follower
\end{itemize}

Before the actual opening of a ticket and the specification of the involved roles, the ticket creator hast to create the ticket in a relatively complex decision process as shown in figure \ref{fig:create}.\\
\begin{figure}
\includegraphics[width=\linewidth]{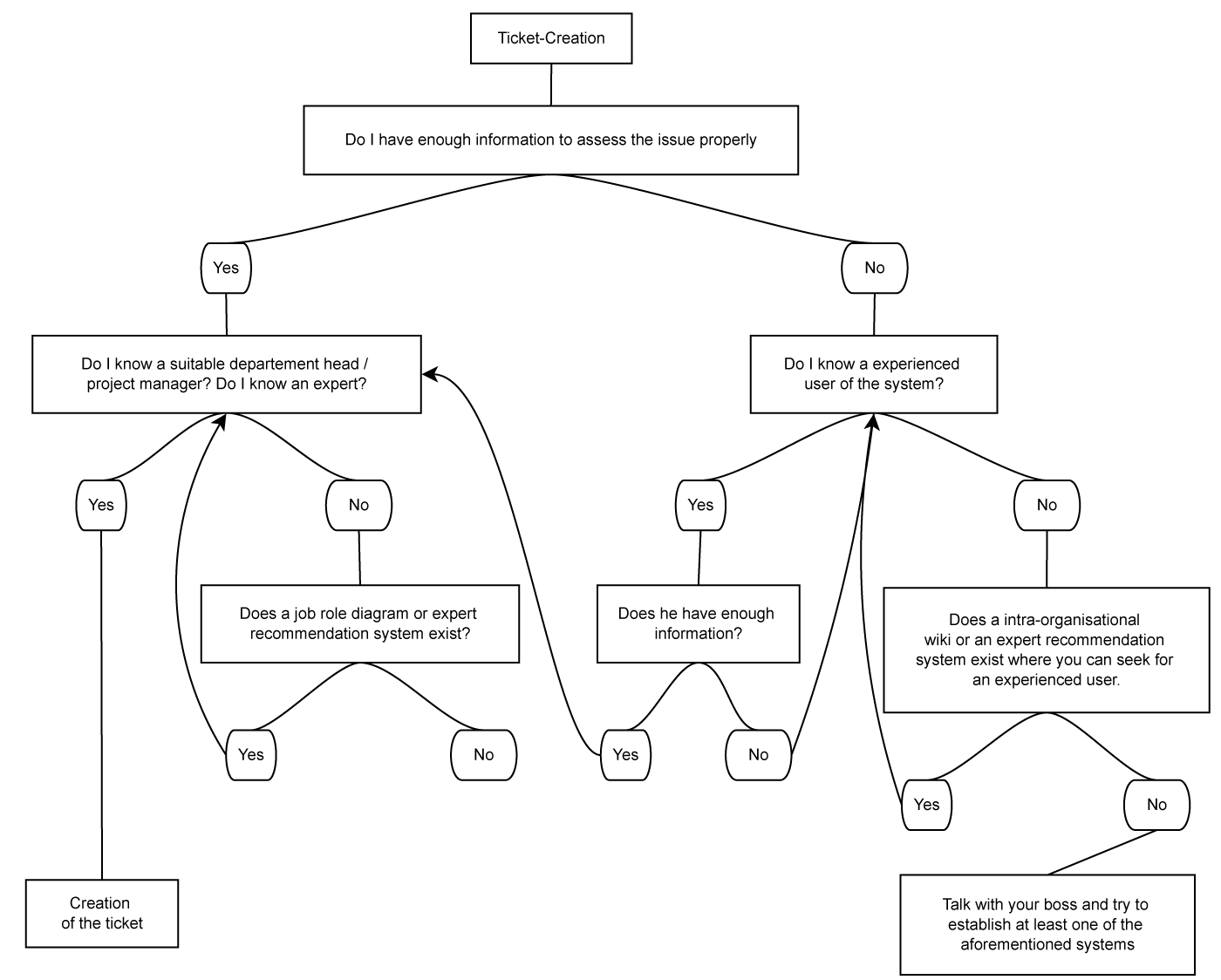}
\caption{Process of Ticket Creation}
\label{fig:create}
\end{figure}
\textbf{Ticket Reporter}
The creator needs to have enough information and knowledge to distinguish if the occurred problem was a problem of misuse (could lead to a feature request) or a bug in the used system.
If the creator can distinguish the problem, it either means that he can act as an expert for this reported part of the system or that he might know somebody who can act as contact person. That means, the more tickets a particular person has for the same area of work the more likely it is that he has knowledge either about the distribution of tasks and skills or about the integration within the system.
\\\\
\textbf{Ticket Assignee}
The ticket reporter and the assignee have different kinds of knowledge in a specific area. During the process of problem solving the assignee creates new artifacts (e.g. source code) and if it's not repetitive work new knowledge especially about specific details of the implementation. \\
Thus the more tickets the person solves in a particularly area the more profound knowledge he should have.
\\\\
It is likely that the words used within the description of a ticket capture the knowledge required to solve new tickets. That means, to find similar tickets and experts who can possibly help, it is required to classify the text in order to discover the knowledge field addressed by the ticket.

\subsection{Retrieval and Recommendation Model}

We apply a simple statistical tf-idf measure to get the relevancy of the words within the ticket description. After taking the most k relevant words of the ticket (where k is normally a value between 5 and 20) it is possible to search for other tickets which contain similar relevant words. The score for each ticket is then
\label{sec:tf-idf}
\[similarity(s, d) = \frac{\sum\limits_{w \in W} \mathrm{tw}(w, s) \cdot \mathrm{tw}(w, d)}{\sqrt{\sum\limits_{w \in W} tw(w, s)^2 \cdot \sum\limits_{w \in W} tw(w, d)^2}}\] (where $W$ is the set of relevant words, $\mathrm{tw}$ is the td-idf weighting function, $s$ is the source ticket and $d$ the ticket of which you want to find out the similarity). The upper part of the division is the sum of the weights multiplied from both tickets. To normalise the results if the contain different amounts of words the upper part is divided by the second norm of each weight. If you look at $W$ as a vector you can transform the upper formula to \[similarity(s,d) = \frac{\mathrm{tW}(W, s) \times \mathrm{tW}(w, d)}{|\mathrm{tW}(W, s)| \cdot |\mathrm{tW}(W, d)|}\]
(where $\mathrm{tW}$ takes the vector of the words W and weights them with the tf-idf measure in regard to the ticket given in the second argument). We can either apply a linear scoring or an inverted exponential score for the retrieval and ranking of similar tickets to a given new ticket.

We further recommend similar tickets which are semantically similar according to a domain ontology representing the expert vocabulary used in the ticket domain. The underlying rational for this approach is, that experts typically use a topic specific vocabulary which is modelled by the domains ontology.  We apply a taxonomic ontology matcher which identifies the tickets' terms as resources from the expert taxonomy. The matcher computes the similarity between two concepts $c1$ and $c2$ based on the taxonomic distance $d(c1,c2)$ between them, which reflects their respective position in the concept hierarchy. The matcher is able to handle multiple inheritance of concepts at the leaf level of a taxonomy.

The concept similarity is defined as: $sim(c1,c2) = 1 - d_{c}(c1,c2)$. Every concept in a taxonomy is assigned a milestone value. Since the distance between two given concepts in a hierarchy represents the path over the \emph{closest common parent} (ccp), the distance is calculated as:\\
\begin{center}
$d_{c}(c1,c2)=d_{c}(c1,ccp)+d_{c}(c2,ccp)$\\
$d_{c}(c,ccp)=milestone(ccp)-milestone(c)$
\end{center}
The milestone values of concepts in a taxonomy are calculated either
\begin{itemize}
    \item with a \emph{linear milestone decrease} $milestone(n)=1-[l(n)/l(N)]$, where $l(n)$ is the depth of the node n in the taxonomic hierarchy and $l(N)$ is the deepest hierarchy level, or
    \item with an \emph{exponential milestone decrease} $milestone(n)=0.5 / k^{l(n)}$, where $k$ is a factor greater than 1 indicating the rate at which the milestone values decrease along the hierarchy
\end{itemize}

After the retrieval and ranking of (statistically and semantically) similar tickets, in the next step the experts are identified. In general, the expertise score is a function $a \times t \mapsto e$ for a given author $a$ and a topic $t$. The set of potential experts are the persons involved in these tickets either as creator, solver, or follower. For each person an experts score is calculated as the accumulated similarity measure (linear or inverted exponential). We distinguish two dimensions, the organisation score and the developer score and calculate the overall expert score $e = e_{o} + e_{d}$.

Additionally, we retrieve relevant Wikipages from an Enterprise Wiki documenting expert knowledge. We assume that an individual who contributes content relevant to a specific ticket topic to such a Wiki has expertise in this topic. we calculate a simple contribution based expertise score as follows:

\begin{equation}
  expertise_{simple} (a, t) \mathrel{\mathop:}= \sum \limits_{s \in S_{a,t}} weight_s(level(s)) + \sum \limits_{w \in W_{a,t}} weight_w(w)
\end{equation}

where by $S_{a,t}$ we denominate the set of sections that cover the topic $t$ and under which author $a$ has contributed content. $t$ is again identified as a concept in the tickets' domain ontology. $weight_s$ is a weighting factor depending on the section level. We used a simple milestone metric in order to express the relevance of a section according to its level. The underlying assumption is that a section with a higher level is about a more general topic, and contributions to a highly specialized topic should be reflected with a higher weighting in the expertise score. Accordingly, by $W_{a,t}$ we denominate all occurrences of terms contributed by author $a$ that can be mapped to ontology topic $t$, weighted by a relevance function $weight_w$.

We further semantically consolidate the expertise score by considering authors Wiki contributions about topics which are semantically similar to the topic of the addressed ticket topic.  We utilize the class hierarchy established by the \texttt{owl:sub\-ClassOf} OWL property and other selectable subtypes of \texttt{owl:ObjectProperties} in order to capture concept relatedness. \\
\begin{figure}
\begin{center}
	\includegraphics[width=8cm]{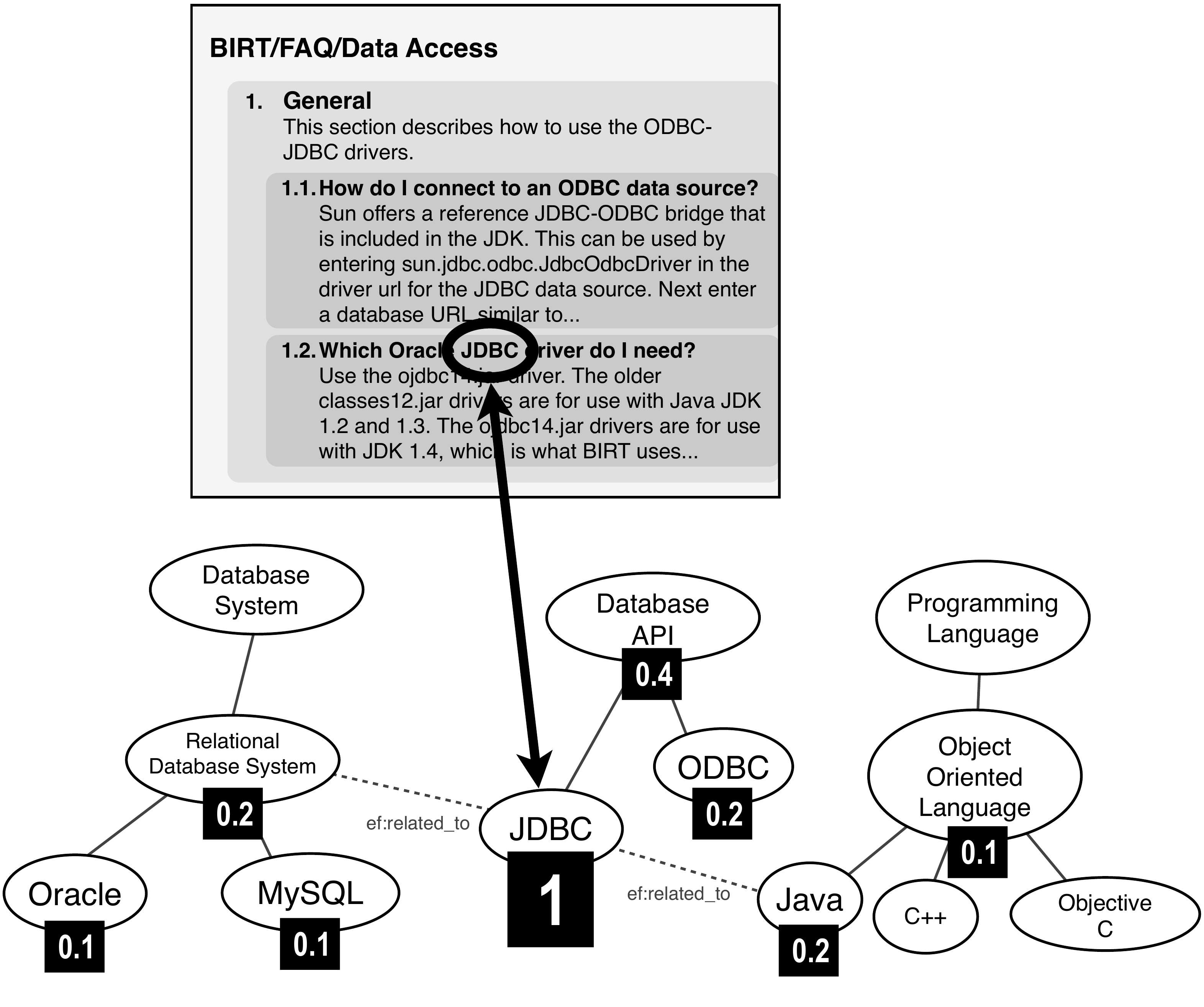}
	\caption{Weighting of Detected and Related Features}
	\label{oc:fig:ef_contributions}
\end{center}
\end{figure}\\
Based on these relations, we calculate a relevance score using ontology based similarity measures until a defined threshold is reached (see figure \ref{oc:fig:ef_contributions}). All concepts with a similarity value higher than this threshold are considered similar and added to the feature set, weighted by its similarity value. This yields a consolidated expertise score which is calculated as follows:

\begin{equation}
  expertise (a, t) \mathrel{\mathop:}= \sum \limits_{t_{sim} \in T, sim(t, t_{sim}) \geq sim_{min}} expertise_{simple}(a,t_{sim}) \cdot sim(t, t_{sim}) .
\end{equation}

Even if topic $t$ is never referenced by author $a$, but neighbouring topics $t_{sim}$ with a similarity to $t$ greater than the threshold $sim_{min}$, then $t$ benefits from $a$'a expertise in each topic $t_{sim}$. The more similar $t$ and $t_{sim}$ are, the higher the benefit for $t$. This score can be further adjusted by additionally considering the authors reputation based on e.g. consindering the revision history, stability, and life time of Wiki entries. \cite{Schafermeier:2011}

\section{Implementation}
\label{implementation}

\subsection{Basic features of the expert recommendation system}
The developed plugin for Jira delivers the following basic features:

\begin{itemize}
\item Automatic indexing of all tickets with usage of

	\begin{itemize}

		\item An adapter for the connection between the Index and the Jira ticket data structure
		\item Automatic translation for ensuring a consistent language basis

	\end{itemize}
	
\item Recommendation of alternative tickets which are similar to the current ticket

	\begin{itemize}

		\item Calculation of different rankings (administrational and development score)
        \item Recommendation of experts

	\end{itemize}
	
\item Integration of matching Wiki articles and adapted expert score from the search engine and expert finder
\end{itemize}

\subsection{System overview}

\begin{figure}
\begin{center}
\includegraphics[width=9cm]{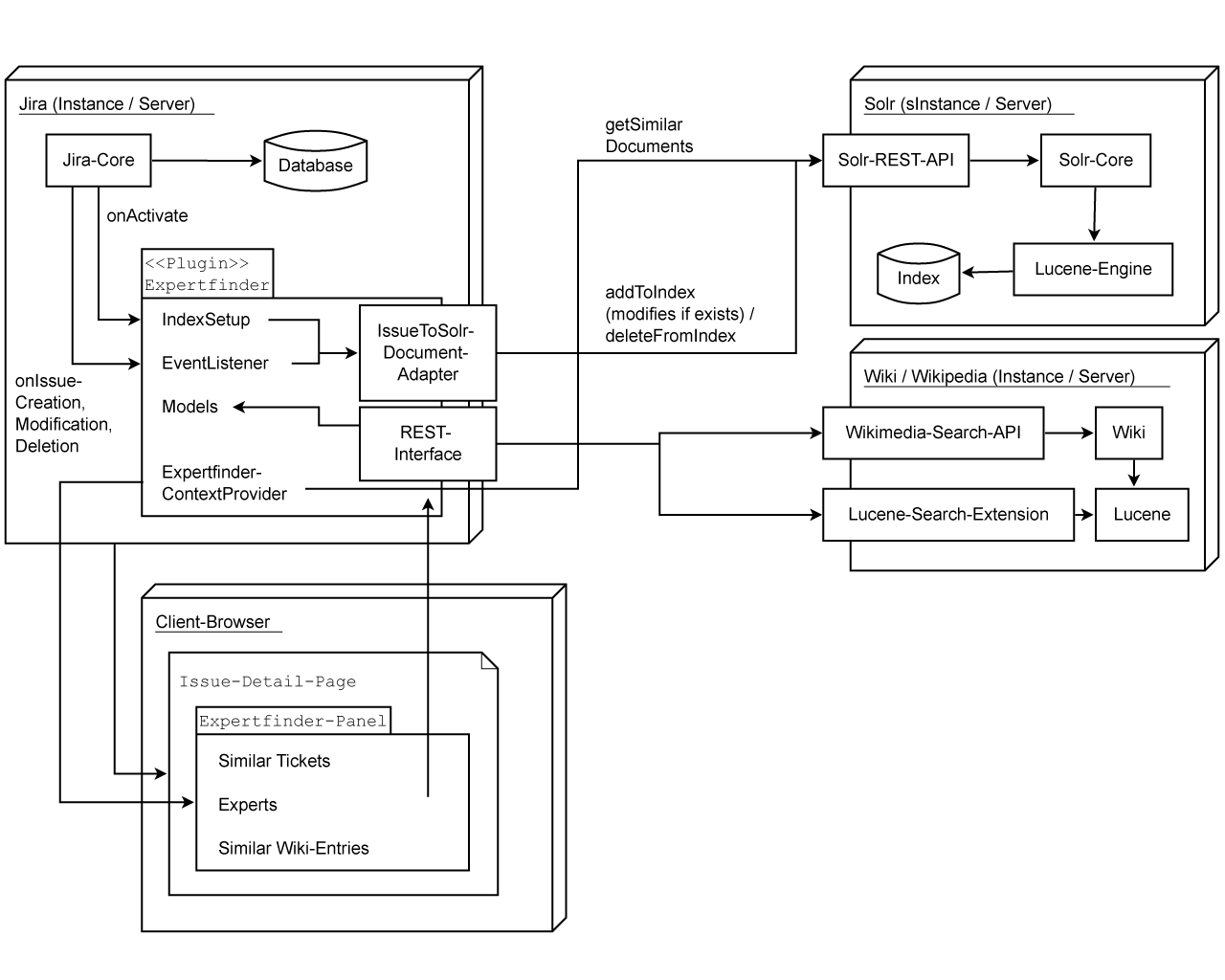}
\end{center}
\caption{Overview of the interaction between the different systems}
\label{fig:overview}
\end{figure}
As indexing and information retrieval platform Lucene with it's efficient indexing engine is used. On top a Solr server manages the Lucene index, supplies Lucene with data, and retrieves the data. Solr gives a REST connector, over which all established (CRUD-) operations are realized.
\\
For the semantic similarity computation we use the CSW Semantic Similarity Matchmaking Framework (SemF) \footnote{\url{http://www.corporate-semantic-web.de/technologies.html}}. The framework allows taxonomic and non-taxonomic concept matching techniques to be applied to selected object properties.\\
\\
The Mediawiki software does have its own search engine to use for finding articles but their features are limited. It is not possible to search for more than one term at once otherwise all given words have to appear in the returned articles. With the extension LuceneSearch \footnote{\url{https://www.mediawiki.org/wiki/Extension:Lucene-search}} it is possible to make more complex queries and give a ranking for the returned articles. To calculate the semantic expert score from the Wiki contributions we adapt and integrate the Wiki expert finder described in \cite{Schafermeier:2011}.

Figure \ref{fig:overview} shows the interaction between the different systems and components. Jira is at the core of the system. From Jira the EventListener is activated to handle all operations on a ticket to send them to the Solr instance. The ExpertFinderContextProvider is triggered when a user requests a ticket details page and gathers all necessary information from Solr, processes it and fills the view.

\section{Evaluation}
\label{evaluation}
For the evaluation employees from the IT department of a German midsized company with around 60 employees were asked to give estimates for best fitting experts for a self-chosen ticket they already worked on themselves. Afterwards all experts for the chosen tickets were collected from the system. Altogether 32 tickets have been evaluated, 98 experts have been given and 267 experts where proposed by the system. The results are shown in \ref{tab:precision-recall} and \ref{fig:precision-recall}.\\
\begin{table}
\begin{scriptsize}
	\begin{center}
	\setlength{\tabcolsep}{5pt}
	\renewcommand{\arraystretch}{1.5}
    \begin{tabular}{|l|l|}
    \hline
    82.7\% & recall \\ \hline
    88.6\% & of all experts are under the first ten results \\ \hline
    74.3\% & of all experts are under the first three results \\ \hline
    45.7\% & of all experts are at the first position of the results \\ \hline
    10.03 & possible experts are proposed by the recommender in average \\ \hline
    \end{tabular}
    \caption {System evaluation}
    \end{center}
    \label{tab:precision-recall}
\end{scriptsize}
\end{table}\\
\begin{figure}
\begin{center}
\includegraphics[width=213pt]{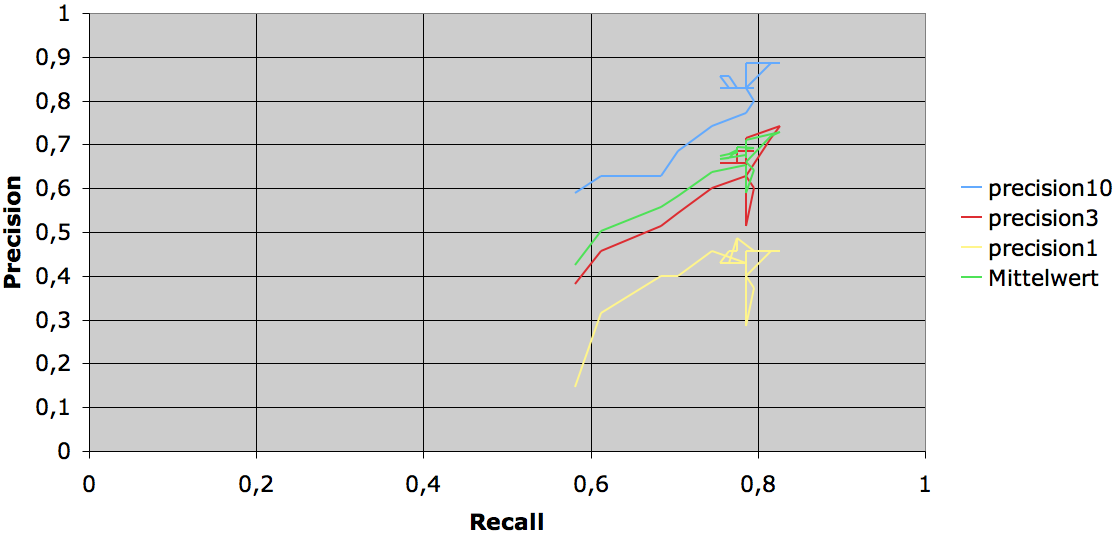}
\end{center}
\caption{Evaluation Precision-Recall diagram}
\label{fig:precision-recall}
\end{figure}\\
As described in section \ref{sec:tf-idf} it is possible to configure how many words should be take into account as relevant for the similarity measurement. This has a direct impact on the quality of the results as shown in figure \ref{fig:bestparam}.\\
\begin{figure}
\begin{center}
\includegraphics[width=213pt]{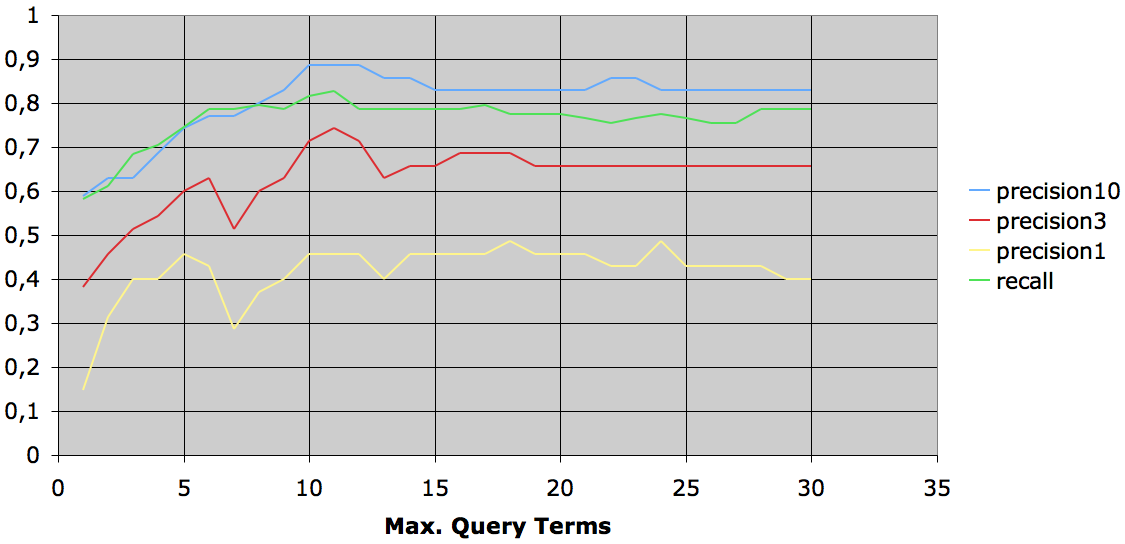}
\end{center}
\caption{Best parameter for the amount of relevant words / terms}
\label{fig:bestparam}
\end{figure}\\
The Wiki extension has been tested and evaluated using the project Wiki of the Eclipse Foundation \footnote{\begin{scriptsize}\url{http://wiki.eclipse.org/Main_Page}\end{scriptsize}} and the Software Engineering Ontology (SEOntology) \cite{Wongthongtham2006}.

\section{Conclusion}
\label{conclusion}
The recommender already reaches relatively good results. Future work might additionally consider multi-lingual translations and wordnets together with larger background knowledge coming, e.g. from linked open data sources such as DBPedia. In future work a comparison on the basis of the the bug tracking data from the eclipse project which was used by some of the related works would be useful. A problem to solve is the import of this data into a Jira bug tracking instance. Furthermore, there are many other information stored or available in the bug tracking system which could be used to optimise the results, e.g. the history of tickets (the change of assignees, the change of the status or of the resolution).

\section{Acknowledgements}
This work has been partially supported by the InnoProfile project "Corporate Semantic Web" funded by the German Federal Ministry of Education and Research (BMBF).

\bibliographystyle{plain}
\bibliography{bibliography}
\end{small}
\end{document}